\documentclass[prl,twocolumn,superscriptaddress,nofootinbib,preprintnumbers]{revtex4-1}

\usepackage{graphicx}
\usepackage{graphics}
\usepackage{amsmath,amssymb}
\usepackage[usenames]{color}
\usepackage{subfigure}
\usepackage{hyperref}
\usepackage{breakurl}
\usepackage{enumitem}
\usepackage{lgreek}

\newcommand{\be}{\begin{equation}}
\newcommand{\ee}{\end{equation}}
\newcommand{\bea}{\begin{eqnarray}}
\newcommand{\eea}{\end{eqnarray}}
\newcommand{\ben}{\begin{enumerate}}
\newcommand{\een}{\end{enumerate}}
\newcommand{\bit}{\begin{itemize}}
\newcommand{\eit}{\end{itemize}}

\newcommand{\la}[1]{\label{#1}}

\newcommand{\Eq}[1]{Eq.~(\ref{#1})}

\newcommand{\Fig}[1]{Fig.~\ref{#1}}

\def\nl{\nonumber \\}

\newcommand{\vv}[1]{\mathbf #1}							% 3-vector
\newcommand{\bert}{\raise-0.45mm\hbox{\Large$\Box$}}			% D'Alembertian

\newcommand{\gd}{\gamma_\downarrow}						% Decay rate
\newcommand{\gu}{\gamma_\uparrow}						% Pumping rate

\makeatletter
\newcommand*\bigcdot{\mathpalette\bigcdot@{.5}}
\newcommand*\bigcdot@[2]{\mathbin{\vcenter{\hbox{\scalebox{#2}{$\m@th#1\bullet$}}}}}
\makeatother

\definecolor{BrickRed}{cmyk}{0,0.89,0.94,0.28}					%%%PANTONE 1805
\definecolor{MidnightBlue}{cmyk}{0.98,0.13,0,0.43}				%%%PANTONE 302
\definecolor{DarkGreen}{rgb}{0.100806,0.495968,0.209979}
\definecolor{orange}{rgb}{0.587167,0.354498,0.146197}

\begin{document}
\preprint{IFT-UAM/CSIC-19-53}

\title{Quantum Theory of Triboelectricity}

\author{Robert Alicki}\email{robert.alicki@ug.edu.pl}
\affiliation{International Centre for Theory of Quantum Technologies (ICTQT), University of Gda\'nsk, 80-308, Gda\'nsk, Poland}
\author{Alejandro Jenkins}\email{alejandro.jenkins@ucr.ac.cr}
\affiliation{International Centre for Theory of Quantum Technologies (ICTQT), University of Gda\'nsk, 80-308, Gda\'nsk, Poland}
\affiliation{Laboratorio de F\'isica Te\'orica y Computacional, Escuela de F\'isica, Universidad de Costa Rica, 11501-2060, San Jos\'e, Costa Rica}
\affiliation{Instituto de F\'isica Te\'orica UAM/CSIC, Cantoblanco, 28049 Madrid, Spain}

\date{First version: 21 Apr.\ 2019. Published in Phys.\ Rev.\ Lett.\ {\bf 125}, 186101 (2020)}

\begin{abstract}
We propose a microphysical theory of the triboelectric effect by which mechanical rubbing separates charges across the interface between two materials.  Surface electrons are treated as an open system coupled to two baths, corresponding to the bulks.  Extending Zel'dovich's theory of bosonic superradiance, we show that motion-induced population inversion can generate an electromotive force.  We argue that this is consistent with the basic phenomenology of triboelectrification and triboluminescence as irreversible processes, and we suggest how to carry out more precise experimental tests.
\end{abstract}

\maketitle

%%%%%%%%%%
%%% INTRODUCTION
%%%%%%%%%%

\section{Introduction}
\la{sec:intro}

The word {\it electricity} comes from the ancient Greek \begin{greek}>'hlektron\end{greek} for amber, a solid material that charges when rubbed with silk or fur.  In the sixth century BCE, pre-Socratic philosopher Thales of Miletus pointed to magnets and amber as evidence of ``a soul or life even to inanimate objects'' \cite{Laertius, Thales}.  The microphysics of dry friction remains poorly understood and there is still no widely accepted theory of triboelectrification, the separation of charges by rubbing.  The Bohr--van Leeuwen theorem establishes that classical physics cannot explain the properties of magnetic materials \cite{VanVleck}, but it is less widely appreciated that classical electrodynamics is insufficient to account for triboelectricity.

Consider the triboelectric generator shown schematically in \Fig{fig:generator}.  The inner cylinder of material $A$ rotates about its axis with angular velocity $\Omega$.  For the right choice of material $B$ in the outer, hollow cylinder, a voltage is established between $A$ and $B$, which can sustain a current $I$ through an external circuit.  The classical electromotive force (emf) $\cal E$ vanishes by the Maxwell-Faraday law:
\be
{\cal E} \equiv \oint \vv E \cdot d \vv s = - \frac{d}{dt} \int \vv B \cdot d \vv a = 0 ,
\ee
as there is no significant variation of the net magnetic flux through the plane of the circuit.  Thus, at the interface between the two materials $A$ and $B$, electrons are being transported {\it against} the average electric field by a nonconservative force (the emf), effectively acting as a negative resistance.  The power for this evidently comes from the motor that spins $A$.  But how mechanical energy is converted into the electrical work done by the emf calls for explanation.

Note that the generation of an emf by the relative motion of $A$ and $B$ must be irreversible, since the direction of the emf cannot depend on the sign of $\Omega$.  On the emf as an active nonconservative force, and on the impossibility of accounting for it using potentials, see \cite{emf}.  Recently, the irreversible dynamics of work extraction by a quantum system coupled to an external disequilibrium has become a subject of theoretical and practical interest in quantum thermodynamics \cite{QT}.

In 1971, Zel'dovich described a process, later dubbed ``superradiance'' by Misner, by which the kinetic energy of a moving dielectric can be partially converted into coherent radiation \cite{Zeldovich1, Zeldovich2}.  This result played a key role in the development of black-hole thermodynamics and it provides a useful guide to a broad class of active, irreversible processes \cite{Bekenstein, BCP}.  As in a laser, superradiance depends on population inversion, which in the case of rotational superradiance results from the disequilibrium associated with the dielectric's macroscopic motion.  Work may then be extracted from the population-inverted states through stimulated emission while generating entropy in the rotating dielectric, which we may treat as a moving heat bath \cite{rotatingbath}.

The exclusion principle prevents stimulated emission of fermions, and therefore their superradiance.  However, we will show here that the motion-induced population inversion of fermions can sustain a macroscopic current between two baths coupled to those fermion states.  Such a process has not, to our knowledge, been considered before, although the authors of \cite{AdS2} noted the presence of Fermi surfaces of singularities in the Green's functions of fermions in the background of a charged black hole.  Here we argue that this offers a plausible theory of triboelectricity, including such remarkable phenomena as the generation of X-rays by peeling ordinary adhesive tape \cite{X-ray1, X-ray2}.

Experimentalists have stressed that triboelectrification and associated effects depend strongly on the relative velocity of the materials in contact and are therefore essentially off equilibrium \cite{Deryagin}.  The process that we describe here is irreversible and velocity dependent.  As such, it is qualitatively different from the reversible processes, describable in terms of Hamiltonians, considered in recently proposed theories of the triboelectric effect \cite{flexo} and the related phenomenon of contact electrification \cite{contact}.  More details on the theoretical and experimental motivations for our non-Hamiltonian, open-system theory of triboelectricity are provided in the Supplemental Material \cite{supplemental}.

\begin{figure} [t] 
\begin{center}
	\includegraphics[width=0.3 \textwidth]{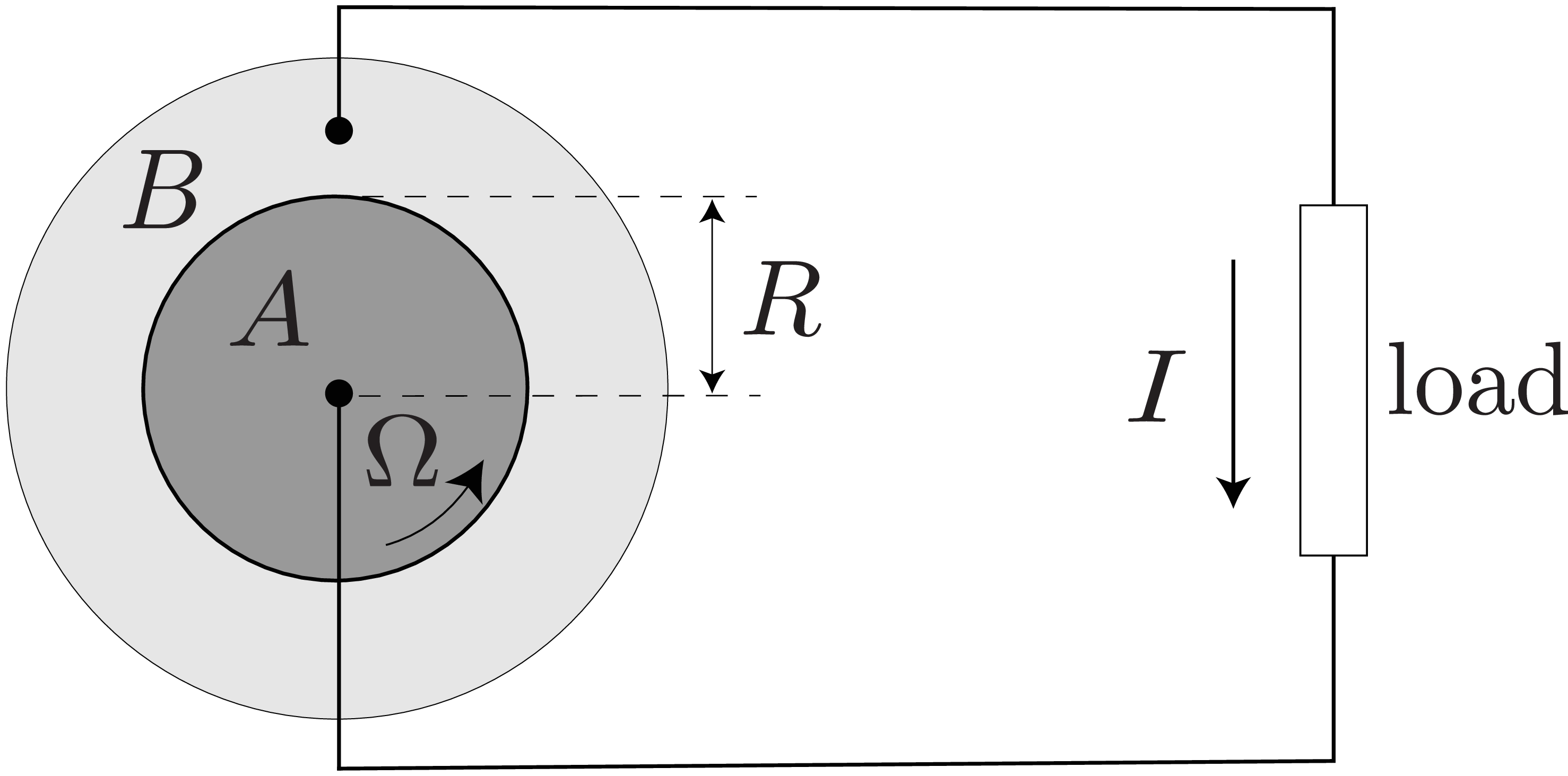}
\end{center}
\caption{\small The triboelectric generator sketched maintains a current $I$ along the circuit if an external motor spins the cylinder of material $A$ and radius $R$ at a sufficient angular velocity $\Omega$ with respect to another material $B$.\la{fig:generator}}
\end{figure}

%%%%%%%%%%
%%% OPEN QUANTUM SYSTEM
%%%%%%%%%%

\section{Open system}
\la{sec:open}

 Consider surface electrons as an open quantum system, weakly coupled to two baths corresponding to bulk materials $A$ and $B$.  In accordance with the setup of \Fig{fig:generator}, we assume cylindrical symmetry so that each electron mode, both in the surface and in the bulk, is labeled by the common magnetic quantum number $m$ (our final results will not, however, depend on this cylindrical symmetry).  Any remaining quantum numbers are labeled by $\sigma$ and $\kappa$.

The second-quantization formalism and notation are similar to those applied to rotational superradiance in \cite{rotatingbath}.  Annihilation and creation operators are, respectively, denoted by $c_{\, \cdot} (\cdot, \cdot)$ and $c^\dagger_{\, \cdot} (\cdot, \cdot)$, while the corresponding energies are denoted by $\omega_{\, \cdot} (\cdot, \cdot)$.  The subsystem is indicated by the index, while the quantum numbers of the mode are given as arguments.  We work in $\hbar = 1$ units.

At rest, the system Hamiltonian is the sum of terms
\be
H_0^x = \sum_{\sigma, m} \omega_x (\sigma, m) c_x^\dagger (\sigma, m) c_x(\sigma, m)
\ee
for $x= a,b$, with $a$ corresponding to the surface attached to material $A$, and $b$ corresponding to the surface attached to material $B$.  Meanwhile, the Hamiltonians for the baths are
\be
H_0^X = \sum_{\kappa, m} \omega_X (\kappa, m) c^\dagger_X (\kappa, m) c_X (\kappa, m)
\ee
for $X = A, B$.

If the material $A$ rotates with an angular velocity $\Omega$ small enough that its internal states are not excited by the rotation, then we have effective Hamiltonians
\be
H_\Omega^a = \sum_{\sigma, m} \big[ \omega_a (\sigma, m) - m \Omega \big] c_a^\dagger (\sigma, m) c_a(\sigma, m)
\la{eq:Haeff}
\ee
and
\be
H_\Omega^A = \sum_{\kappa, m} \big[ \omega_A (\kappa, m) - m \Omega \big] c^\dagger_A (\kappa, m) c_A (\kappa, m)
\la{eq:HAeff}
\ee
The sign of $\Omega$ in Eqs.\ (\ref{eq:Haeff}) and (\ref{eq:HAeff}) is arbitrary and has been chosen for later convenience. The shift from the $H_0$'s to the $H_\Omega$'s may be interpreted as a Doppler shift.

The experimental evidence is now clear that triboelectrification of solids is dominated by electron tunneling processes \cite{Wang}. We therefore consider a weak interaction between the surface electrons and each of the two baths,
\be
H_X^x = \sum_{\kappa, \sigma, m} g_X^x (\kappa, \sigma, m) c^\dagger_X (\kappa, m) c_x(\sigma, m) + \hbox{h.c.} ,
\la{eq:HXx}
\ee
where the $g_X^x$'s correspond to direct transition amplitudes, to which the Coulomb interaction probably contributes significantly.

We expect the surface states $a$ and $b$ to be localized along the transport direction (i.e., perpendicular to the surface), so that their mutual interaction plays no role in transport.  We therefore neglect $ab$ interactions, which would give only a hybridization absorbable into modified wave functions.  Moreover, since the $ab$ interaction is not needed to obtain a triboelectric effect, it is reasonable to neglect it for the sake of simplicity since our present goal is to formulate a qualitatively new model rather than a detailed one.  We therefore take the full Hamiltonian to be
\be
H_{\rm full} = H_\Omega^a + H_0^b+ H_\Omega^A + H_0^B + H_A^a + H_B^a + H_A^b + H_B^b .
\la{eq:Hfull}
\ee

%%%%%%%%%%
%%% KINETIC EQUATIONS
%%%%%%%%%%

\section{Kinetic equations}
\la{sec:kinetic}

The occupation numbers for the surface electron states are:
\be
n_x (\sigma, m) = \left\langle c_x^\dagger (\sigma, m) c_x(\sigma, m) \right\rangle .
\ee
In the limit of weak coupling between the system and the baths, we may compute the decay rates $\gd^{xX}$ using Fermi's golden rule \cite{RA-Fermi, AL}.  The pumping rates $\gu^{xX}$ are related to the decay rates by the Kubo-Martin-Schwinger (KMS) condition.  Omitting the quantum numbers, the corresponding kinetic equation may be written as
\be
\dot n_x = \gu^{xA} + \gu^{xB} - \left(  \gd^{xA} + \gd^{xB} + \gu^{xA} + \gu^{xB} \right) n_x .
\la{eq:dna}
\ee
Let us define
\be
n_X(y) \equiv \frac{1}{e^{\beta(y - \mu_X)} + 1} ,
\ee
where $\mu_X$ is the chemical potential of the corresponding bulk material in equilibrium.

By Fermi's golden rule, the rate of decay of the $a$ surface electrons into the bath $A$ is
\bea
\gd^{aA} (\sigma, m) = 2 \pi \big[ 1 - n_A( \omega_a (\sigma, m) ) \big] \overline{g_A^a}^2(\sigma, m) ,
\la{eq:aAd}
\eea
where
\be
\overline{g_A^a}^2(\sigma, m) \equiv \sum_\kappa \left| g_A^a (\kappa, \sigma, m) \right|^2
\delta \big( \omega_a (\sigma, m) - \omega_A(\kappa, m) \big) .
\ee
For the pumping rate we have, by the KMS condition,
\bea
\gu^{aA}(\sigma, m) &=& 2 \pi n_A \big( \omega_a(\sigma, m) \big) \overline{g_A^a}^2(\sigma, m) \nl
&=& e^{-\beta(\omega_a(\sigma, m) - \mu_A)} \gd^{aA}(\sigma, m) .
\la{eq:aAu}
\eea

Because of the shift of the energies in \Eq{eq:Haeff}, for the rate of decay of $a$ surface electrons into the bath $B$ we have
\be
\gd^{aB} (\sigma, m) = 2 \pi \big[ 1 - n_B( \omega_a (\sigma, m) - m\Omega ) \big] \overline{g_B^a}^2(\sigma, m; \Omega) ,
\la{eq:aBd}
\ee
where
\bea
\hskip -0.4 cm \overline{g_B^a}^2(\sigma, m; \Omega) \!\! &\equiv& \!\! \sum_{\kappa'} \left| g_B^a (\kappa', \sigma, m) \right|^2 \nl
&& \times \delta \big(\omega_a (\sigma, m) - m \Omega - \omega_B(\kappa', m) \big) .
\eea
The pumping rate is given by the modified KMS relation
\be
\gu^{aB}(\sigma, m) = e^{-\beta(\omega_a(\sigma, m) - m \Omega - \mu_B)} \gd^{aB}(\sigma, m) .
\la{eq:aBu}
\ee
Thus, when
\be
m \Omega > \omega_a(\sigma,m) - \mu_B
\la{eq:aDs}
\ee
the corresponding state exhibits population inversion ($\gu^{aB} > \gd^{aB}$), making it possible to extract electrical work from it.  A similar analysis gives us $\gd^{bX}$ and $\gu^{bX}$.  Equation (\ref{eq:aDs}) corresponds to the ``anomalous Doppler shift'' of the Ginzburg-Frank theory of radiation by uniformly moving sources \cite{Ginzburg1, Ginzburg2}.

Work may be extracted by superradiance from a single moving bath because the pumping of the population-inverted bosonic state leads to stimulated emission \cite{rotatingbath}.  In the case of fermions, on the other hand, a second bath is needed to remove the pumped fermion from its population-inverted state, before another fermion becomes available to sustain an active current.  Whereas superradiance and other forms of bosonic radiation by uniformly moving charges may be described classically \cite{Bekenstein, Ginzburg1}, the fermionic case (which we propose here as the microphysical basis of the triboelectric effect) requires a quantum treatment.

%%%%%%%%%%
%%% TRIBOCURRENTS
%%%%%%%%%%

\section{Tribocurrents}
\la{sec:currents}

In the steady state ($\dot n_a = 0$), \Eq{eq:dna} implies that
\be
n_a = \bar n_a \equiv \left( \gu^{aA} + \gu^{aB} \right) / \Gamma^a ,
\ee
where
\be
\Gamma^a \equiv \gu^{aA} + \gd^{aA} + \gu^{aB} + \gd^{aB} .
\ee
For each channel $(\sigma, m)$, the number of electrons per unit time that flow from $A$ to $a$ is
\be
j_a = \gu^{aA} - \left( \gd^{aA} + \gu^{aA} \right) \bar n_a.
\la{eq:ja1}
\ee
By Eqs.\ (\ref{eq:aAu}) and (\ref{eq:aBu}), this can be reexpressed as
\be
j_a = \gu^{aA}\gd^{aB} \left[ 1 - e^{\beta (m \Omega + \mu_B - \mu_A)} \right] / \Gamma^a .
\la{eq:ja2}
\ee
In the steady state this is also the current the flows from $B$ to $a$ (see \Fig{fig:currents}).

Similarly, $\dot n_b = 0$ implies that
\be
n_b = \bar n_b \equiv \left( \gu^{bA} + \gu^{bB} \right) / \Gamma^b ,
\ee
where
\be
\Gamma^b \equiv \gu^{bA} + \gd^{bA} + \gu^{bB} + \gd^{bB} .
\ee
The current that flows from $B$ to $b$ (which in the steady state equals the current from $b$ to $A$) is then
\bea
j_b &=& \gu^{bB} -\left( \gd^{bB} + \gu^{bB} \right) \bar n_b \nl
&=& \gd^{bA}\gu^{bB} \left[1 - e^{-\beta (m \Omega + \mu_B - \mu_A)} \right] / \Gamma^b .
\la{eq:jb} 
\eea
As illustrated in \Fig{fig:currents}, the total electric current from $A$ to $B$ is
\be
J = - e \left[ \sum_{\sigma, m} j_a (\sigma, m) - \sum_{\sigma',m} j_b (\sigma',m) \right] .
\la{eq:J}
\ee

By Eqs.\ (\ref{eq:aAu}) and (\ref{eq:aBd}) we have that
\be
\gu^{aA}\gd^{aB} \sim n_A \big( \omega_a(\sigma, m) \big) \big[ 1 - n_B \big( \omega_a (\sigma, m) - m\Omega \big) \big] .
\ee
As the ratio $\mu/k_B T$ for ambient temperature is \hbox{$\simeq 10^{2}$}, we replace the Fermi-Dirac distributions by step functions, $n_X(y) \simeq H(\mu_X - y)$, giving
\be
\gu^{aA}\gd^{aB} \sim  \chi_{[\mu_B + m\Omega, \, \mu_A]} \big( \omega_a(\sigma, m) \big) ,
\ee
where $\chi_E $ is the indicator function of the set $E$.  Thus, only surface modes of electrons satisfying
\be
m \Omega  < \mu_A - \mu_B
\la{eq:cond-a}
\ee 
contribute to the tribocurrent $j_a$ in \Eq{eq:ja2}, so that $j_a > 0$.  By a similar reasoning we find that only modes satisfying
\be
m \Omega > \mu_A - \mu_B
\la{eq:cond-b}
\ee
contribute to $j_b$ in \Eq{eq:jb} and therefore $j_b > 0$.

\begin{figure} [t] 
\begin{center}
	\includegraphics[width=0.25 \textwidth]{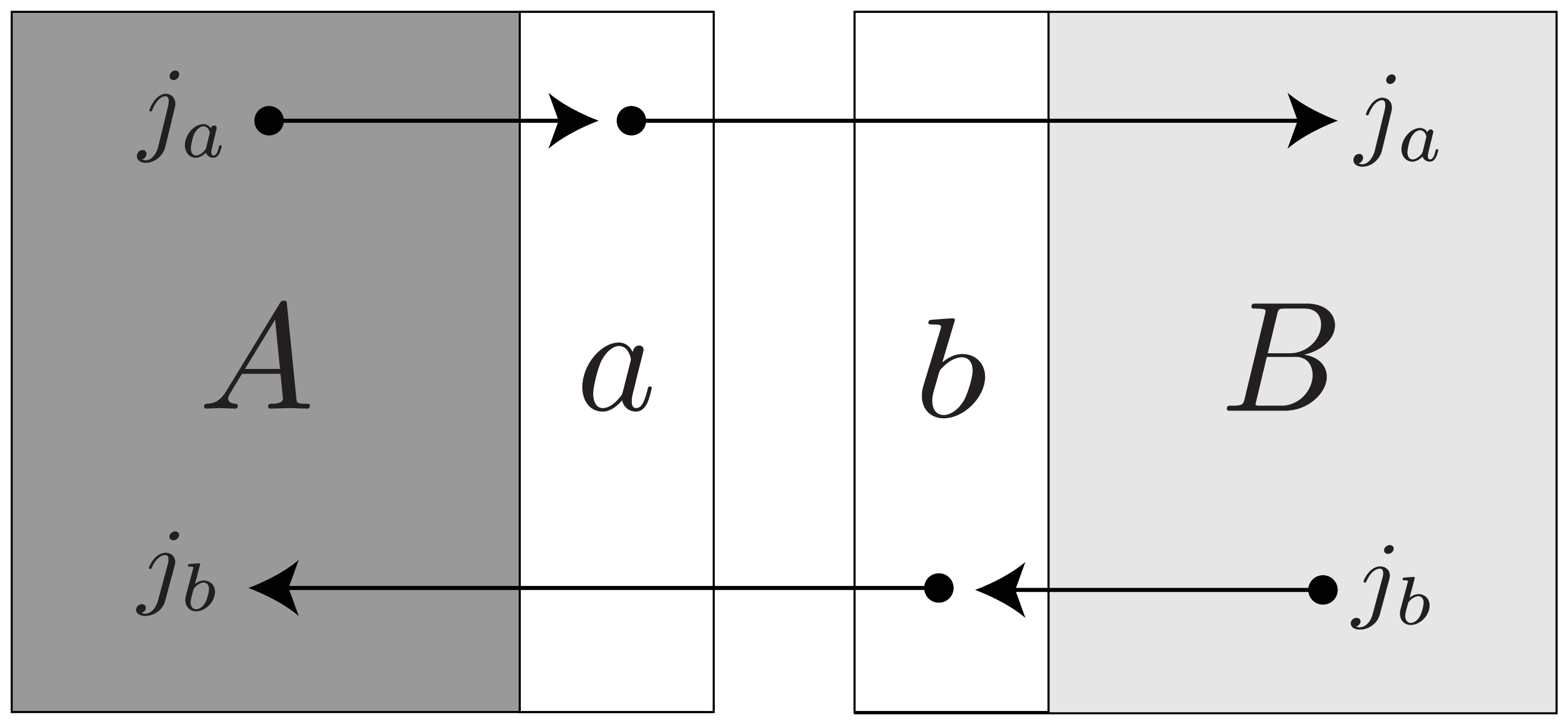}
\end{center}
\caption{\small Sketch of the currents $j_a$ of \Eq{eq:ja1} and $j_b$ of \Eq{eq:jb}, for the open system $a,b$ in a steady state.\la{fig:currents}}
\end{figure}

%%%%%%%%%%
%%% PHENOMENOLOGY
%%%%%%%%%%

\section{Phenomenology}
\la{sec:pheno}

The $j_x$ currents depend on surface-to-bulk tunneling rates that are exponentially sensitive to potential barrier heights and widths.  The sums in \Eq{eq:J} also depend on the density of surface electron states.  But even without detailed characterization of this complex landscape we can show that our theory is nontrivially consistent with key observations.

The sign of $J$ in \Eq{eq:J} depends on the relative magnitudes of $\gu^{aA}\gd^{aB} / \Gamma^a$ and $\gd^{bA}\gu^{bB} / \Gamma^b$, controlled by the couplings between bulks and surfaces.  For two materials in rubbing contact, the sign of $J$ can therefore vary with the surface's geometry, corrugation, stress, etc.  This agrees with the observation of patches of positive and negative charge, with sizes at the roughness scale $\simeq 1 ~ \mu {\rm m}$ \cite{mosaic}.

According to Eqs.\ (\ref{eq:cond-a}) and (\ref{eq:cond-b}), as $| \mu_A - \mu_B |$ increases under net charging, fewer modes contribute to the $j_x$ in \Fig{fig:currents} giving the charging, while more modes contribute to the opposing current.  This may explain why significant triboelectrification is usually seen only when two materials well separated in the ``triboelectric series'' are rubbed against each other \cite{series1}.  It may also explain why the net current between the rubber belt and the metal brush is opposite at the two terminals of a Van de Graaff generator, where the brushes are identical except for their respective voltages \cite{Purcell}.

A larger work function for material $A$ implies a higher barrier for $a$ to $B$ tunneling, thus suppressing $\gd^{aB}$ in \Eq{eq:ja2}, whereas a larger work function for material $B$ suppresses $\gd^{bA}$ in \Eq{eq:jb}.  We therefore expect net $J$ (for zero initial voltage) to tend to point from the material with greater work function to the one with smaller work function, as reported in \cite{series2}.  Work functions do not, however, determine triboelectric properties entirely. The details of the interface barrier can play an important role, especially for insulators \cite{Wu}.

Let $(k_z , k_m)$ be the cylindrical components of the wave vector and let $k_F$ be the maximum value of $\sqrt{k_z^2 + k_m^2}$, corresponding to the Fermi wave vector for the surface electrons.  In terms of the linear speed \hbox{$V_s = | \Omega R |$} with which the surface of material $A$ slides against the surface of material $B$ in \Fig{fig:generator},
\be
| m \Omega | = | k_m V_s | \leq k_F V_s .
\ee
From Eqs.\ (\ref{eq:cond-a}) and (\ref{eq:cond-b}) we conclude that
\be
e \phi_{\rm oc} = \left| \mu_A - \mu_B \right|_{\rm at~ zero~ current} \lesssim \hbar k_F V_s ,
\la{eq:maxV}
\ee
where $\phi_{\rm oc}$ is the tribovoltage (note that we have reintroduced $\hbar$).  The bound of \Eq{eq:maxV} is saturated if and only if $j_a$ is negligible compared to $j_b$, or vice versa.

Taking \hbox{$k_F \simeq 1 ~\hbox{\AA}^{-1}$} and \hbox{$V_s \simeq 1$ m/s} in \Eq{eq:maxV}, we obtain \hbox{$\phi_{\rm oc} \lesssim 10^{-5}$ V}.  Rapid mechanical separation of the charged surfaces increases the voltage accordingly \cite{contact}.  If the distance between the charged surfaces grows from angstrom to meter scale, the resulting voltage will be \hbox{$\lesssim 10^5$ V}, as in a Van de Graaff generator \cite{Purcell}.  If the distance goes from interatomic to \hbox{$\simeq 10 ~\mu$m} scale, the energy of the electrons can be in the visible range (\hbox{$\simeq 1$ eV}).  On triboluminescence, see \cite{luminescence} and references therein.

The surface charge density generated by peeling adhesive tape increases strongly with the peel rate \cite{Deryagin}.  The surface charge density $\simeq 10^{10} \, e / \hbox{cm}^2$ reported in \cite{X-ray2} may be consistent with our theory, supposing that the maximum velocity of slippage between the dissimilar materials in contact is larger, by a couple of orders of magnitude, than the average peel rate $\simeq 1$ cm/s.  The X-ray bursts produced by the peeling are preceded by a further hundredfold increase in the charge density, in a process connected with macroscopic stick-slip oscillations \cite{X-ray2}.  Such acoustic oscillations can enhance the effective $m \Omega$ in the exponential of \Eq{eq:aBu}, pumping the $\phi_{\rm oc}$ by another 2 or 3 orders of magnitude.

Recent experiments, in which various materials are charged using a uniform technique, find triboelectric charge densities $\sigma$ lying on an approximately symmetric interval $[-\sigma_{\rm max}, \, \sigma_{\rm max}]$; see Fig.~3 in \cite{series2}.  Since the maximum and minimum values of $\sigma$ correspond to entirely different materials, this symmetry has no obvious explanation in potential models.  On the other hand, it agrees with our \Eq{eq:maxV}, according to which $\sigma_{\rm max}$ (proportional to the upper bound on $\phi_{\rm oc}$) should be determined by the technique used.  More detailed comparison to data will require a better understanding of how the effective $V_s$ depends on the various experimental setups.

%%%%%%%%%%
%%% DISCUSSION
%%%%%%%%%%

\section{Discussion}
\la{sec:discussion}

 Ginzburg stressed that ``radiation during the uniform motion of various sources is a universal phenomenon rather than an eccentricity'' \cite{Ginzburg2}, with counterparts ``in any field theory'' \cite{Ginzburg1}.  Considering bosonic superradiance in terms of open quantum systems clarifies the respective roles of macroscopic motion, dissipation, and stimulated emission \cite{rotatingbath}.  Here we have extended that analysis to fermions, allowing us to propose a microphysical explanation of the persistent conversion of macroscopic motion into an emf, something that cannot be obtained from density functional theory or other equilibrium descriptions \cite{luminescence}.

In our treatment, the emf results from motion-induced enhancement of pumping over decay (i.e., population inversion) in the modified KMS relation of \Eq{eq:aBu}.  This allows us to obtain active currents from the kinetic equations for the populations of the surface electron states coupled to the two bulk materials.  This theory has other key features qualitatively different from what one might expect in a potential description: Rubbing produces opposing currents $j_a$ and $j_b$ (see \Fig{fig:currents}), and the upper bound on charging of \Eq{eq:maxV} (approached when the two materials are very far from each other on the triboelectric series) depends only on the Fermi wave vector of the surface electrons and on sliding velocity.  We have argued that these and other aspects of our theory are compatible with reported observations.  New experiments with precise control of the sliding velocity (possibly based on setups closer to \Fig{fig:generator}) could test our predictions more directly.

Some authors have interpreted triboelectrification as resulting from phonon production by mechanical rubbing \cite{PanZhang}.  The irreversible consumption of mechanical power by dry friction may result from the generation of phonons that then thermalize in the bulk \cite{FL}. Such phonons may contribute to the tribocurrent by assisting electron tunneling, enhancing the effective $g_X^x$'s in \Eq{eq:HXx}.  On the other hand, the direct $j_x$'s consume power even when dry friction is not accompanied by significant net charging. The roles of phonons and $j_x$ currents in both dry friction and triboelectrification therefore call for further investigation. In the Supplemental Material we sketch an argument for why we expect the contribution of phonon-assisted tunneling to triboelectrification to be relatively small. \\

%%%%%%%%%%
%%% ACKNOWLEDGMENTS
%%%%%%%%%%

\begin{acknowledgments} {\bf Acknowledgments:}  We thank Carlos D\'iaz, Jos\'e Gracia-Bond\'ia, Gian Guzm\'an-Verri, Itamar Kimchi, Lok Lew Yan Voon, John McGreevy, and Niclas Westerberg for discussions.  R.\ A.\ was supported by the International Research Agendas Programme (IRAP) of the Foundation for Polish Science (FNP), with structural funds from the European Union (EU).  A.\ J.\ was supported by the University of Costa Rica's Vice-rectorate for Research (Project No.\ 112-B6-509), by the EU's Horizon 2020 research and innovation program under the Marie Sk{\l}odowska-Curie Grant No.\ 690575, and by the Polish National Agency for Academic Exchange (NAWA)'s Ulam Programme (Project No.\ PPN/ULM/2019/1/00284, ``Energy conversion by open quantum systems: Theory and applications'').  \end{acknowledgments}

%%%%%%%%%%
%%% SUPPLEMENTAL MATERIAL
%%%%%%%%%%

\bibliographystyle{h-physrev}
\bibliography{ref}
\clearpage
\newcommand{\beginsupplement}{%
        \setcounter{table}{0}
        \renewcommand{\thetable}{S\arabic{table}}%
        \setcounter{figure}{0}
        \renewcommand{\thefigure}{S\arabic{figure}}%
        \setcounter{equation}{0}
        \renewcommand{\theequation}{S\arabic{equation}}%
     }
\onecolumngrid
\beginsupplement

\section{Supplemental Material}

Triboelectrification is a complex, multi-scale phenomenon involving atomic/molecular as well as collective, macroscopic processes \cite{luminescence}.  Moreover, it depends not only on material properties but also on surface roughness, stresses, environmental conditions, and other details of the experimental setup \cite{contact}.  Despite the history of triboelectric experiments stretching back to antiquity and the importance of the subject in modern material science and technology, the microphysics of triboelectricity and associated processes was poorly understood and hotly debated until recently (see \cite{X-ray2} and references therein).

Experiments with Kelvin probe force microscopy have now established that the triboelectrification of solids is associated primarily with the transfer of electrons across the interface of dissimilar materials in contact (see \cite{Wang} and references therein), forming a charged double layer at that interface.  This has corroborated the interpretation put forth in the 20th century by the Soviet school of Deryagin et al.\ \cite{Deryagin}.  Recent theoretical work has described this physics in terms of tunneling between electronic surface states and bulk modes described by a tight-binding Hamiltonian \cite{contact}.

Deryagin et al.\ also stressed that triboelectrification is strongly dependent on the speed with which the materials in contact move relative to each other.  This is demonstrated, e.g., by the marked increase in the work required to peel off adhesive tape as the rate of the peeling is increased.  They interpreted this as proof that triboelectrification is an essentially off-equilibrium process, not describable in terms of potentials \cite{Deryagin}.  More recently, Collins et al.\ has stressed that ``the very presence of triboluminescence demonstrates
that charges are being rearranged in a non-reversible manner'' and that ``it would be na\"ive to try and impose an equilibrium theory [e.g., Density Functional Theory (DFT)] to a necessarily non-equilibrium process.'' \cite{luminescence}

An even more fundamental ---though less widely appreciated--- consideration is that the electromotive force (emf) is the integral of an active, nonconservative force that can do work on a charge going around in a closed path and can pump a charge up along an existing potential \cite{emf}.  This implies that no microphysical explanation of the triboelectric emf can be obtained from a purely Hamiltonian description.  The theoretical treatments of contact electrification and the triboelectric effect that have been published previously can describe the transfer of electrons between dissimilar materials in contact, but they cannot account for the emf.

This puts triboelectricity in the category of {\it active processes} that irreversibly extract work from an underlying thermodynamic disequilibrium.  The dynamics of such processes has been a blind spot of theoretical physics, which for the most part has been framed to treat only conservative or passively irreversible processes, or the responses of passive systems to an external driving.  An adequate dynamical description of this work extraction requires a treatment in terms of open systems with positive feedback.  In a quantum-mechanical context, this has recently become an active subject of investigation in ``quantum thermodynamics'' \cite{QT}, but relatively little work in that area has been done so far for quantum fields (i.e., in second quantization).

Zel'dovich's theory of rotational superradiance \cite{Zeldovich1, Zeldovich2, Bekenstein, BCP} and the Ginzburg-Frank theory of radiation by uniformly moving sources \cite{Ginzburg1, Ginzburg2, Bekenstein, BCP} describe work extraction by a field (in the form of non-thermal radiation) from an underlying disequilibrium induced by macroscopic motion.  However, those descriptions apply to bosonic fields only.  In this work we extend to fermions the open-system treatment of superradiance as an active process that we published previously in \cite{rotatingbath}.  This brings a new class of systems into the purview of quantum thermodynamics and should therefore be of interest beyond the question of triboelectricity, and even beyond materials science.

As in superradiance, the bath's macroscopic motion modifies the KMS relation for the fermionic field (see \Eq{eq:aBu} in the main text).  This causes population inversion of the low-energy fermion states (equivalently, it gives them a negative ``local temperature''), enhancing pumping over decay rates and making it possible to extract work from those states.  The Pauli exclusion principle prevents work from being extracted by the fermions from a single moving bath, because there is no stimulated emission (and therefore no positive feedback) for the population-inverted states \cite{rotatingbath}.  But our analysis in the main text, based on the kinetic equations for the fermion population numbers, establishes that active triboelectric currents can be sustained due to the presence of two different bulk materials, treated as separate baths coupled to the electron surface states.  These currents can be seen as reflecting the presence of a motion-induced emf.

Our treatment of the triboelectric effect is based on the model, standard in solid-state physics, of independent electrons moving in the averaged effective potential that includes a screened Coulomb interaction and which leads to the band structure of the electronic states.  The modification of this effective potential in the vicinity of the surface is accounted for by the introduction of distinct surface states and by the bulk-surface interaction Hamiltonians of \Eq{eq:HXx} in the main text.  The Coulomb interaction probably plays a significant role in the transition between bulk and surface states.  In the present work we have not attempted to compute the $g_X^x$ amplitudes of \Eq{eq:HXx} in any detailed model of the complex landscape of the material surfaces.  Note, however, that our understanding of these amplitudes and of the physics involved in their computation is essentially consistent with the picture invoked in \cite{contact, Wang} and other recent literature on contact electrification.

In order to render tractable an analytic treatment, we make some simplifying assumptions in the description of the triboelectric system.  These simplifications include the cylindrical symmetry assumed in the system illustrated in the main text's \Fig{fig:generator}, the uniformity of the angular velocity $\Omega$, and the assumption that the interaction between the two surfaces in contact gives only a hybridization absorbable into modified wave functions (see Eqs.\ \eqref{eq:HXx} and \eqref{eq:Hfull} in the main text).  These simplifications are analogous, e.g., to assuming periodic boundary conditions or a perfectly regular lattice in solid-state theory.  We argue in the Letter that our main conclusions do not depend on these simplifying assumptions.

Another issue is the role of phonon-assisted tunneling in triboelectrification.  Rubbing of solids must be accompanied by the production of soft phonons whose rapid thermalization accounts for the mechanical power consumed by dry friction \cite{FL}.  Some authors have suggested that these phonons may assist the transport of electrons from one material to the other \cite{PanZhang}.  The generation of the phonons should be described by a dynamical mechanism similar to superradiance, but such a theoretical treatment has not, to our knowledge, been carried out.

\begin{figure} [t] 
\begin{center}
	\includegraphics[width=0.15 \textwidth]{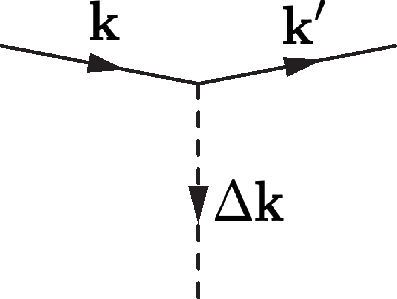}
\end{center}
\caption{\small An electron (solid line) emits or absorbs a phonon (dashed line).\la{fig:phonon}}
\end{figure}

On the other hand, an order-of-magnitude estimation suggests that the contribution of these phonons to the tribocurrents is probably small, and that it is reasonable to ignore it at the present level of complexity in our treatment.  Consider the emission or absorption of a phonon by an electron with initial momentum $\vv k$ and final momentum $\vv k'$, as shown in \Fig{fig:phonon}.  The phonon has momentum
\be
\Delta \vv k = \vv k - \vv k'
\ee
and the change in the energy $E_{\rm el}$ of the non-relativistic electron is therefore
\be
\Delta E_{\rm el} = \frac{\hbar^2}{m} \vv k \cdot \Delta \vv k .
\ee
The energy of the phonon is
\be
E_{\rm ph} = \hbar v_{\rm ph} \Delta k ,
\ee
where $v_{\rm ph}$ is the relevant speed of sound.  The ratio of the phonon energy to the change in the energy of the electron is therefore
\be
\frac{E_{\rm ph}}{\Delta E_{\rm el}} \sim \frac{v_{\rm ph}}{v_{\rm F}} \sim 10^{-3} ,
\ee
where $v_{\rm F}$ is the Fermi velocity $\sim 10^6$ m/s.  This leads us to expect that phonons will not have a large effect on the behavior of the surface electrons.  However, the respective roles of the phonons and the tribocurrents, in both pure dry friction and triboelectrification, certainly call for further investigation.

In the Letter's main text we have pointed out various significant ways in which our theory is consistent with the basic phenomenology of the triboelectric effect.  Some of these are, in our view, highly nontrivial, including the approximate symmetry of the interval $[-\sigma_{\rm max}, \, \sigma_{\rm max}]$ for the charge densities reported in \cite{series2}.  Some of the relevant experiments cited (see \cite{Wang, series2}) were published after our theory first appeared as a preprint in the physics arXiv, in April of 2019.

We expect that a more detailed comparison of our theory to experiment will depend less on the calculation of the $g_X^x$ amplitudes and of the contributions from phonon-assisted tunneling than on better experimental control over the sliding velocity $V_s$ in the main text's \Eq{eq:maxV}.  Though obviously impractical as a triboelectric generator due to the smallness of the voltage produced, the arrangement shown in the Letter's \Fig{fig:generator}, if implemented with precision control over the speed $\Omega$, could provide detailed testing of our theoretical predictions.  This may also help to advance our understanding of the related problem of the microphysics of dry friction.

\end{document}